\documentclass{aastex63}

\usepackage{times}
\usepackage{amsmath,amssymb,amstext}
\hypersetup{linkcolor=red,citecolor=blue,filecolor=cyan,urlcolor=blue}
\usepackage{tabularx}
\usepackage{longtable}
\usepackage{float}
\usepackage{natbib}
\usepackage{graphicx,color}
\usepackage{txfonts}

\usepackage{multirow}
\usepackage{array}
\usepackage{rotating}
\usepackage{sidecap}
\usepackage{hyperref}
\usepackage{epstopdf}
\usepackage{footnote}
\usepackage{tabularx}
\usepackage{booktabs}
\usepackage{longtable}
\usepackage{tabu}
\usepackage{longtable}

\newcommand{\kms}{km\,s$^{-1}$}

\begin{document}

\title{{\large {New Evidence on the Origin of Solar Wind Microstreams/Switchbacks}}}
%\title{{\large {New Evidence on the Coronal Origin of Solar Wind Microstreams/Switchbacks}}}

\author{Pankaj Kumar\altaffiliation{1,2}}
\affiliation{Department of Physics, American University, Washington, DC 20016, USA}
\affiliation{Heliophysics Science Division, NASA Goddard Space Flight Center, Greenbelt, MD, 20771, USA}

\author{Judith T.\ Karpen}
\affiliation{Heliophysics Science Division, NASA Goddard Space Flight Center, Greenbelt, MD, 20771, USA}

\author{Vadim M.\ Uritsky\altaffiliation{3,1}}
\affiliation{Catholic University of America, 620 Michigan Avenue NE, Washington, DC 20064, USA}
\affiliation{Heliophysics Science Division, NASA Goddard Space Flight Center, Greenbelt, MD, 20771, USA}

\author{Craig E.\ Deforest}
\affiliation{Southwest Research Institute, 1050 Walnut Street, Suite 300, Boulder, CO 80302, USA}

\author{Nour E.\ Raouafi}
\affiliation{The John Hopkins University Applied Physics Laboratory, Laurel, MD 20723, USA}

\author{C.\ Richard DeVore}
\affiliation{Heliophysics Science Division, NASA Goddard Space Flight Center, Greenbelt, MD, 20771, USA}

\author{Spiro K.\ Antiochos}
\affiliation{CLaSP, University of Michigan, Ann Arbor, MI, 48109, USA }

\email{pankaj.kumar@nasa.gov}

%*****************************************************************************
\begin{abstract}
Microstreams are fluctuations in the solar wind speed and density associated with polarity-reversing folds in the magnetic field (also denoted switchbacks).  
Despite their long heritage, the origin of these microstreams/switchbacks remains poorly understood.  
For the first time, we investigated periodicities in microstreams during Parker Solar Probe (PSP) Encounter 10 to understand their origin. 
Our analysis was focused on the inbound corotation interval on 2021 November 19-21, while the spacecraft dove toward a small area within a coronal hole (CH). 
Solar Dynamics Observatory remote-sensing observations provide rich context for understanding the PSP in-situ data. 
Extreme ultraviolet images from the Atmospheric Imaging Assembly reveal numerous recurrent jets occurring within the region that was magnetically connected to PSP during intervals that contained microstreams. 
The periods derived from the fluctuating radial velocities in the microstreams (approximately 3, 5, 10, and 20 minutes) are consistent with the periods measured in the emission intensity of the jetlets at the base of the CH plumes, as well as in larger coronal jets and in the plume fine structures. 
Helioseismic and Magnetic Imager magnetograms reveal the presence of myriad embedded bipoles, which are known sources of reconnection-driven jets on all scales.  
Simultaneous enhancements in the PSP proton flux and ionic ($^3$He, $^4$He, Fe, O) composition during the microstreams further support the connection with jetlets and jets. 
In keeping with prior observational and numerical studies of impulsive coronal activity, we conclude that quasiperiodic jets generated by interchange/breakout reconnection at CH bright points and plume bases are the most likely sources of the microstreams/switchbacks observed in the solar wind. 

\end{abstract}
%*****************************************************************************

%*****************************************************************************

%%%%%% Section 1 %%%%%%%%%%%%%%%%%%%%%%%%%%%%%%%%%%%%%%%%%%%%%%%%%%%%%%%%%%

\section{INTRODUCTION}\label{intro}
It is well known that the fast solar wind ($>$500 \kms) originates from polar coronal holes \citep{viall2020}. Microstreams are the fluctuations in the fast wind radial speed  of $\>$20 \kms above or below the running-average speed, lasting 0.4 days \citep{neugebauer1995,neugebauer2021}. These structures have been detected by Helios at 0.3 AU and Ulysses between $\approx$1 and 3 AU \citep{thieme1990,neugebauer1995}. Microstreams are often associated with folds in the magnetic field denoted ``switchbacks", which have been detected by the Parker Solar Probe (PSP; \citealt{fox2016,Raouafi_2022PhT,raouafi_parker_2023}) in the near-Sun solar wind observations \citep{bale2019,kasper2019,horbury2020}. Two theories have been proposed to explain the switchbacks in the solar wind: (i) interchange reconnection in the solar corona \citep{fisk2020,zank2020,drake2021} (ii) in-situ processes in the solar wind \citep{squire2020,ruffolo2020,shoda2021}. PSP observations of the switchbacks suggest a correspondence with the typical spatial scale of granules ($\sim$1000 km), and a correspondence between the switchback patches and the characteristic size of supergranules ($\sim$30,000 km) \citep{bale2021,fargette2021}.

Coronal holes are populated with omnipresent bright points and plumes \citep{deforest2001a,deforest1997,deforest2007,wang2020}.
With the advent of high spatial and temporal resolution from SDO/AIA, tiny jets (so-called ``jetlets") were found to be associated with ``plume transient bright points" occurring at the base of a plume \citep{raouafi2014}. This study supported earlier speculation \citep{raouafi2008} that jetlets are collectively responsible for the enhanced mass and maintenance of plumes, but the dynamic relationship between jetlets and plumes was not understood until recently.  We discovered 3, 5, and 10 minute periods in small, bright EUV bursts and associated outflows (jetlets) at the base of plumes \citep{kumar2022}. Continuing upward, the plumes are composed of numerous ``plumelets" accounting for most of the emission, which are spatially connected to the jetlets and vary on the same timescales  \citep{uritsky2021}. We speculated that p-mode waves possibly play a role in triggering oscillatory reconnection (e.g., \citealt{chen2006}) during the main phase of the plume, and/or modulate the jetlets \citep{kumar2022}.

The origin of microstreams in CHs is still under debate. It has been suggested that microstreams could be associated with CH jets \citep{neugebauer1995, neugebauer2012, sterling2020a,neugebauer2021}, but jets originating from large coronal bright points (e.g., \citealt{sterling2020a, kumar2018}) are not frequent enough to account for the quasiperiodic velocity spikes that characterize the microstreams detected by PSP. Therefore big jets are unlikely to be the microstream sources. In contrast, jetlets at the base of plumes and tiny bright points scattered throughout CHs are much more frequent \citep{raouafi2014,kumar2019a,kumar2022,raouafi2023}, offering a possible quasiperiodic link to microstreams/switchbacks (for convenience we will refer to this phenomenon as microstreams throughout the remainder of this paper). 

A natural question arises here: what constitutes compelling evidence that the patches of solar wind microstreams are due to jetlets?  The observed characteristics should be consistent with these features originating in the low corona and subsequently propagated outward into the solar wind, and consistent with the signatures of interchange reconnection. Consequently, we propose that jetlets and microstreams must have similar variability. To test this idea, we analyzed PSP encounter 10 observations and cotemporal SDO/AIA data when PSP was connected to an equatorial CH. Specifically, we extracted jetlet frequencies at the base of a plume in the CH and compared them with the frequencies of solar wind microstreams. Based on microstream observations during encounter 10, \citet{bale2022} suggested that the fast solar wind is powered by interchange reconnection in CHs. However, a direct connection between dynamic solar activity in the CH and the observed microstreams was not established. Here we report microstream periodicities during the three-day interval of PSP encounter 10 (2021 November 19-21). We have used wavelet analyses of radial wind velocity and EUV intensity to determine the periods in microstreams and jetlets magnetically connected to PSP, respectively, during the same intervals. We find remarkable agreement between the periods measured at the Sun and at 15-30 R$_\sun$, as well as compositional signatures at PSP during the microstream patches that are characteristic of coronal jets. These results strongly indicate a causal relationship between jetlets at the Sun and microstreams in the outer corona and heliosphere.  In \S 2, we present the observations and analysis, in \S 3 we discuss the physical interpretation of the observed periods, and summarize the results in \S4.

%%%%%%%%%%%%%%%%%%%%%%%%%%%%%%%%%%%%%%%%%%%%%%%%%%%%%%%%%%%%%%%

%%%%%%%%%%%%%%%%%%%%%%%%%%%%%%%%%%%%%%%%%%%%%%%%%%%%%%%%%%%%%%
\begin{figure*}
\centering
\includegraphics[height=0.9\textheight]{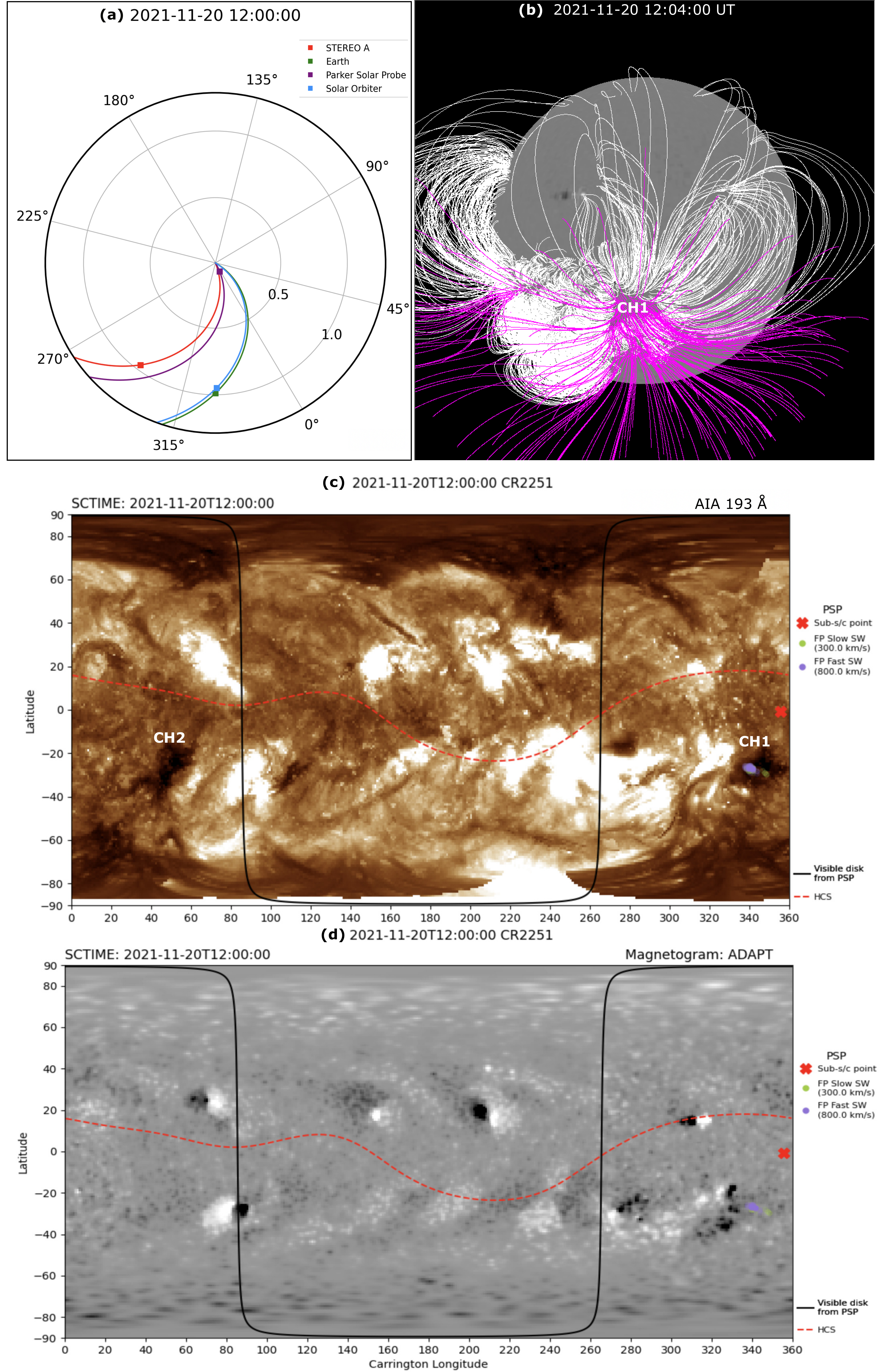}
\caption{(a) PSP position on 2021 November 20 12:00 UT. (b) PFSS extrapolation on November 20, 12:04 UT. White (pink) color marks closed (open) field lines. (c,d) PSP footpoint connectivity to CH1 (AIA 193 {\AA} and photospheric magnetogram synoptic map) on November 20 12:00 UT. PSP was connected to CH1 on November 19-20 and to CH2 on November 21. }
\label{fig1}
\end{figure*}
%%%%%%%%%%%%%%%%%%%%%%%%%%%%%%%%%%%%%%%%%%%%%%%%%%%%%%%%%%%%%%%
%%%%%%%%%%%%%%%%%%%%%%%%%%%%%%%%%%%%%%%%%%%%%%%%%%%%%%%%%%%%%%
\begin{figure*}
\centering
\includegraphics[height=0.35\textheight]{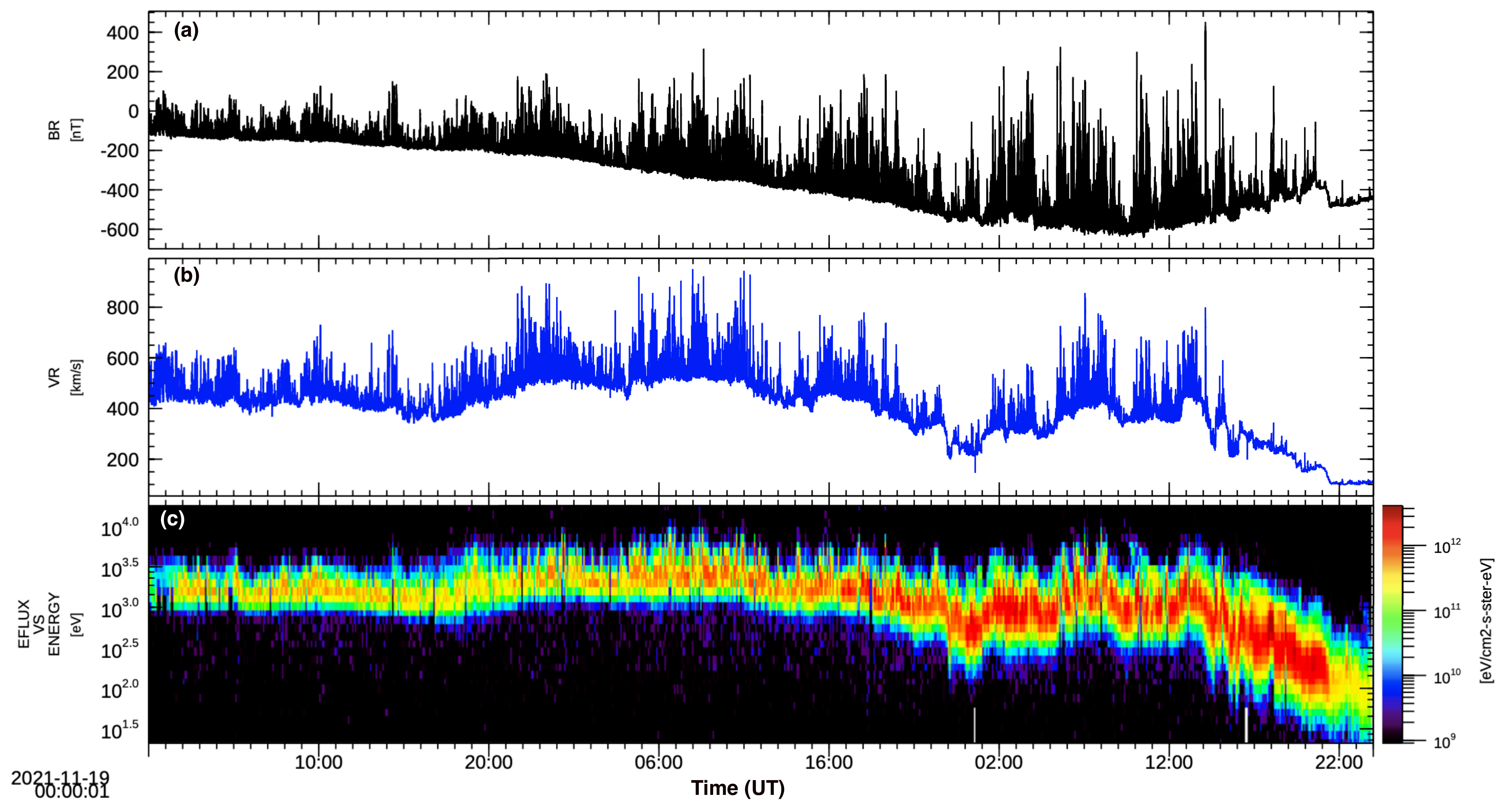}
\includegraphics[height=0.65\textheight]{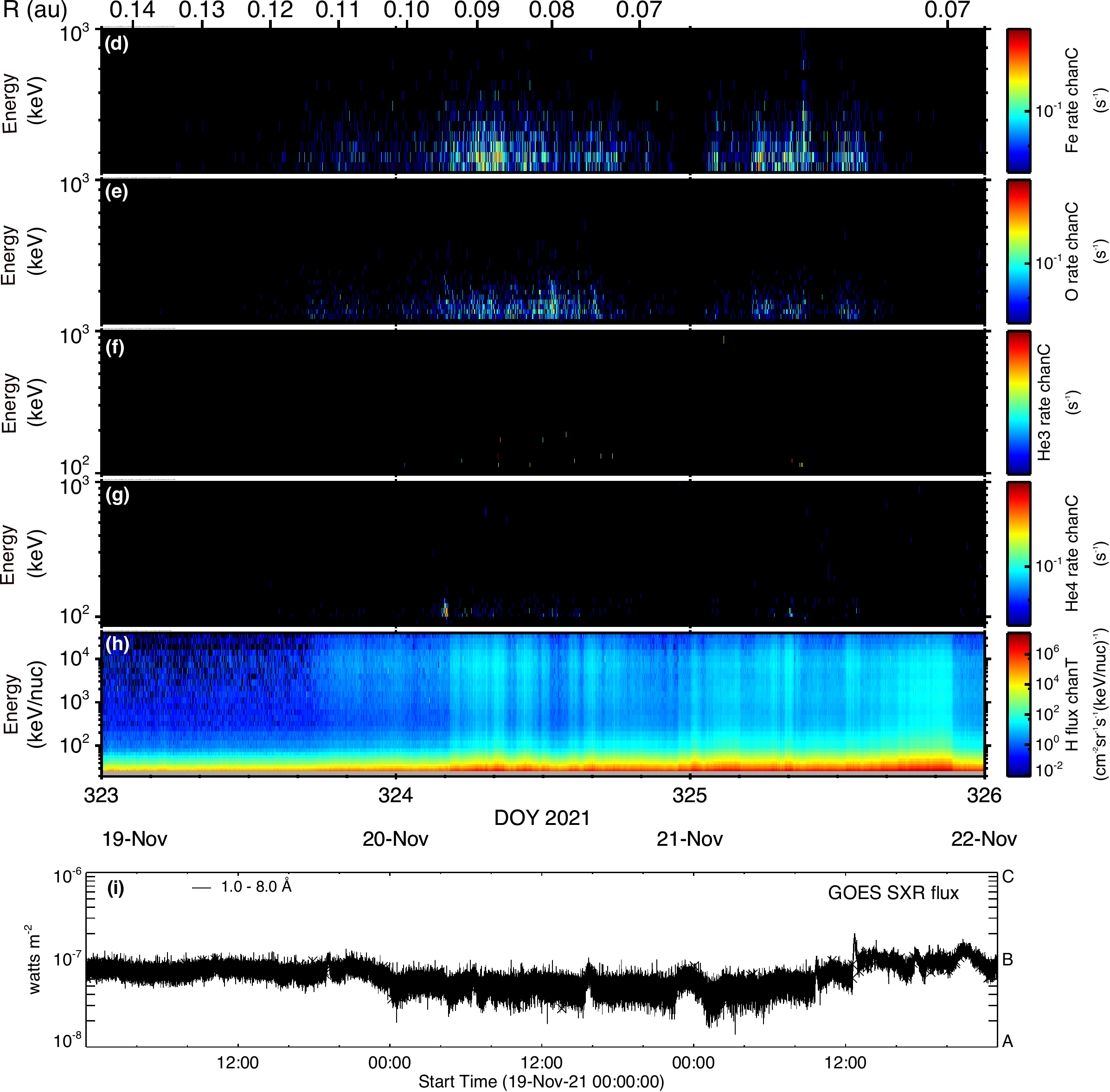}
\caption{(a,b) Solar wind radial magnetic field (B$_R$) and speed (V$_R$) during 2021 November 19-21. (c) Proton energy flux vs energy. (d,e,f,g,h) The top axis indicates the PSP solar radial distance (AU). Ion composition during November 19-21. Spectrograms for iron (Fe), oxygen (O), helium ($^3$He, $^4$He), and protons (H). (i) GOES soft X-ray flux profile (1-8 {\AA}, 1-s cadence) during November 19-21. } 
\label{fig2}
\end{figure*}
%%%%%%%%%%%%%%%%%%%%%%%%%%%%%%%%%%%%%%%%%%%%%%%%%%%%%%%%%%%%%%%

%%%%%%%%%%%%%%%%%%%%%%%%%%%%%%%%%%%%%%%%%%%%%%%%%%%%%%%%%%%%%%
\begin{figure*}
\centering
\includegraphics[height=0.45\textheight]{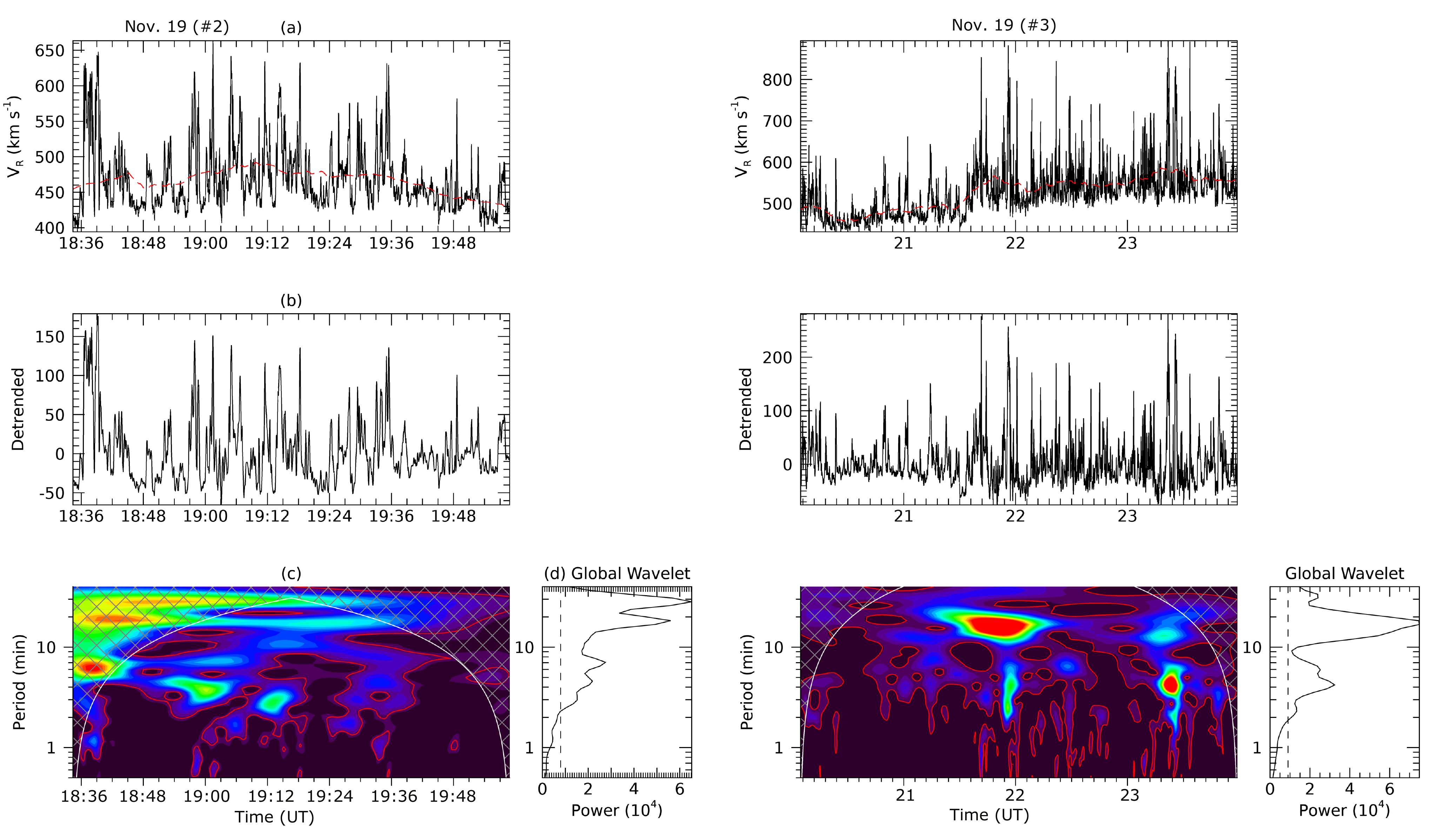}
\includegraphics[height=0.45\textheight]{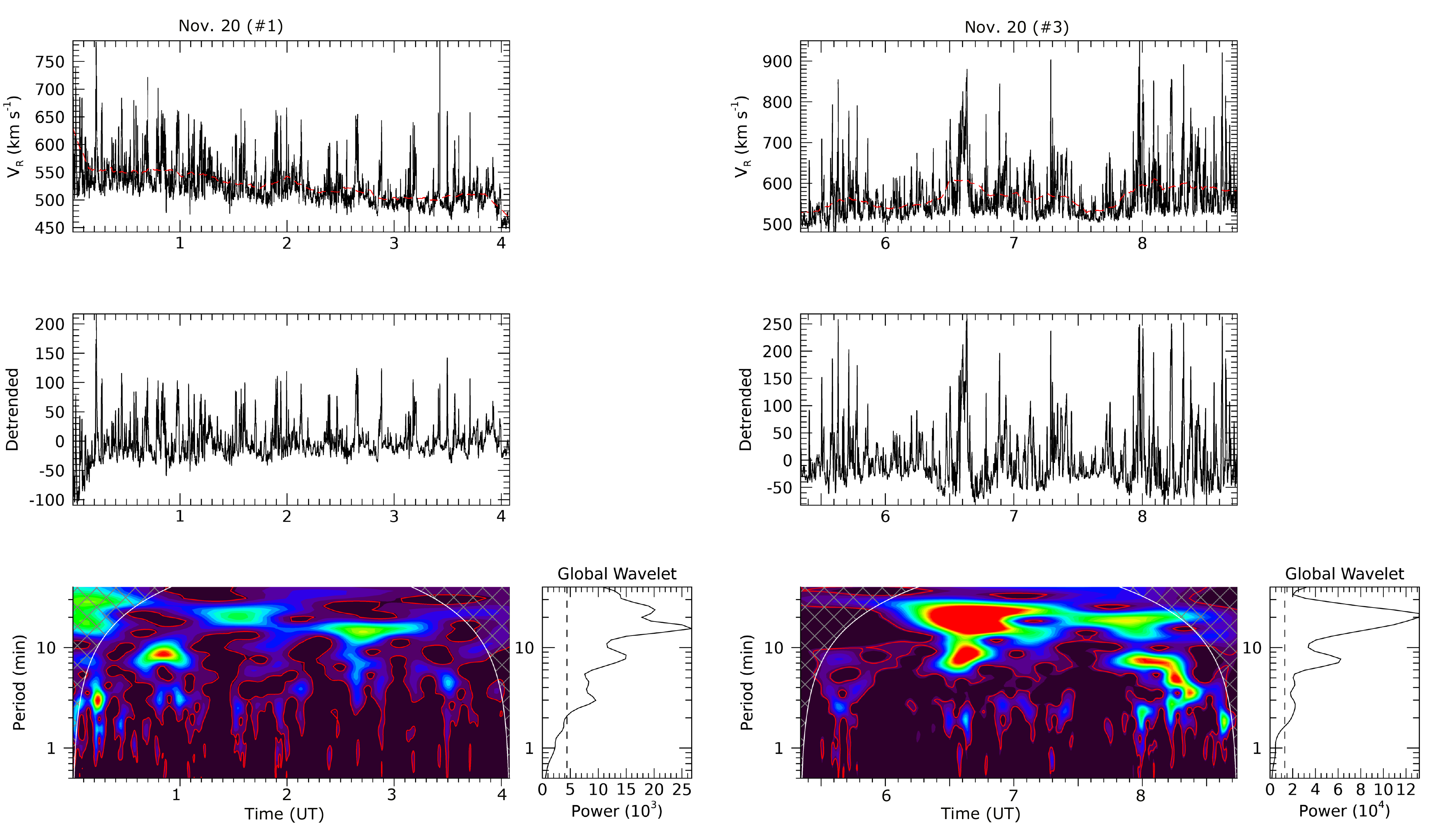}
\caption{Microstream periodicities in four intervals on November 19, 20. {\it Left:} (a) Solar wind radial speed (V$_R$) in Interval \#2. (b) The smoothed and detrended light curve after subtracting the red trend shown in (a) from the original V$_R$. (c) Wavelet power spectrum of the detrended signal. Red contours outline the 99\% significance level. (d) Global wavelet power spectrum. The dashed line is the 99\% global confidence level. {\it Right}: Same methods as in the left panels for Interval \#3.} 
\label{fig3}
\end{figure*}
%%%%%%%%%%%%%%%%%%%%%%%%%%%%%%%%%%%%%%%%%%%%%%%%%%%%%%%%%%%%%%%

%%%%%%%%%%%%%%%%%%%%%%%%%%%%%%%%%%%%%%%%%%%%%%%%%%%%%%%%%%%%%%
\begin{figure*}
\centering{
\includegraphics[width=18cm]{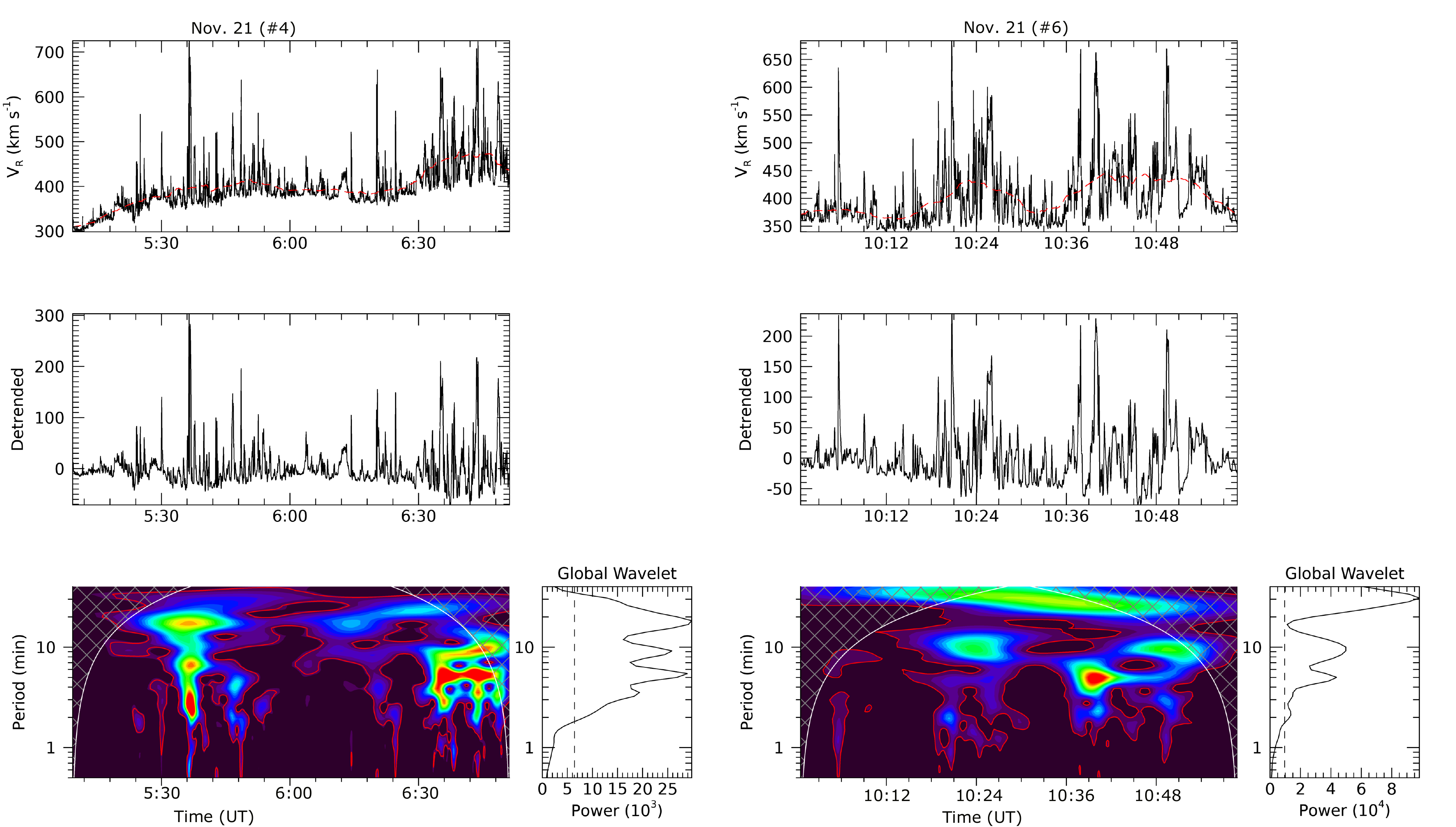}
}
\caption{Microstream periodicities in two intervals on November 21. Same methods as in Figure \ref{fig3}). Left: Interval \#4. Right: Interval \#6.}
\label{fig4}
\end{figure*}
%%%%%%%%%%%%%%%%%%%%%%%%%%%%%%%%%%%%%%%%%%%%%%%%%%%%%%%%%%%%%%%
%%%%%%%%%%%%%%%%%%%%%%%%%%%%%%%%%%%%%%%%%%%%%%%%%%%%%%%%%%%%%%
\begin{figure*}
\centering{
\includegraphics[width=14.3cm]{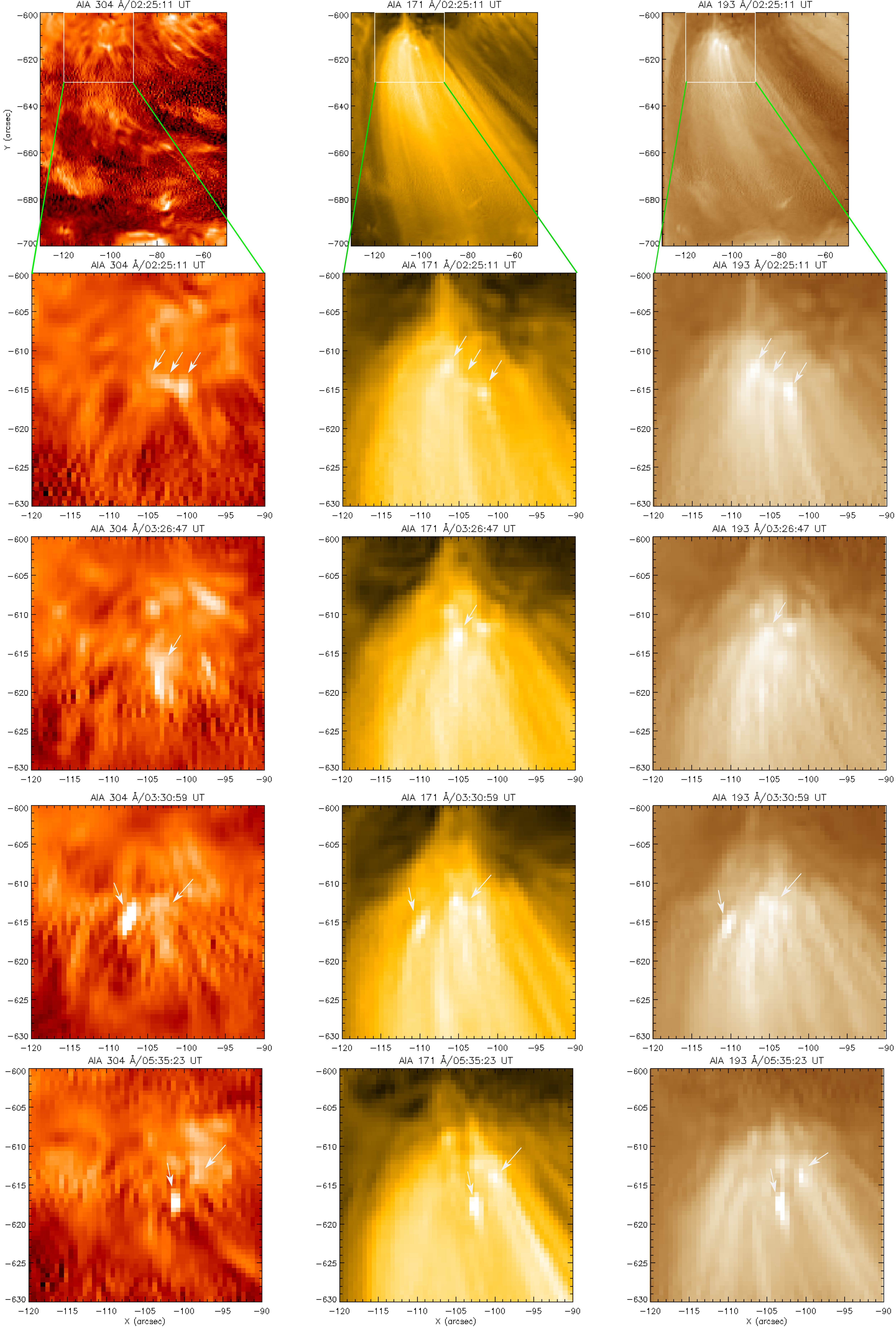}
}
\caption{Jetlets at the base of a plume in CH1 on 2021 November 20. Top row: AIA 304, 171, and 193 {\AA} images of the plume. Remaining rows: zoomed views of the base in the same AIA channels. Arrows point to the footpoint brightenings and associated jetlets propagating along plumelets. An animation (from 01:01:35 UT to 11:57:23 UT) of this figure is available.} 
\label{fig5}
\end{figure*}
%%%%%%%%%%%%%%%%%%%%%%%%%%%%%%%%%%%%%%%%%%%%%%%%%%%%%%%%%%%%%%%

%%%%%%%%%%%%%%%%%%%%%%%%%%%%%%%%%%%%%%%%%%%%%%%%%%%%%%%%%%%%%%
\begin{figure*}
\centering{
\includegraphics[width=16cm]{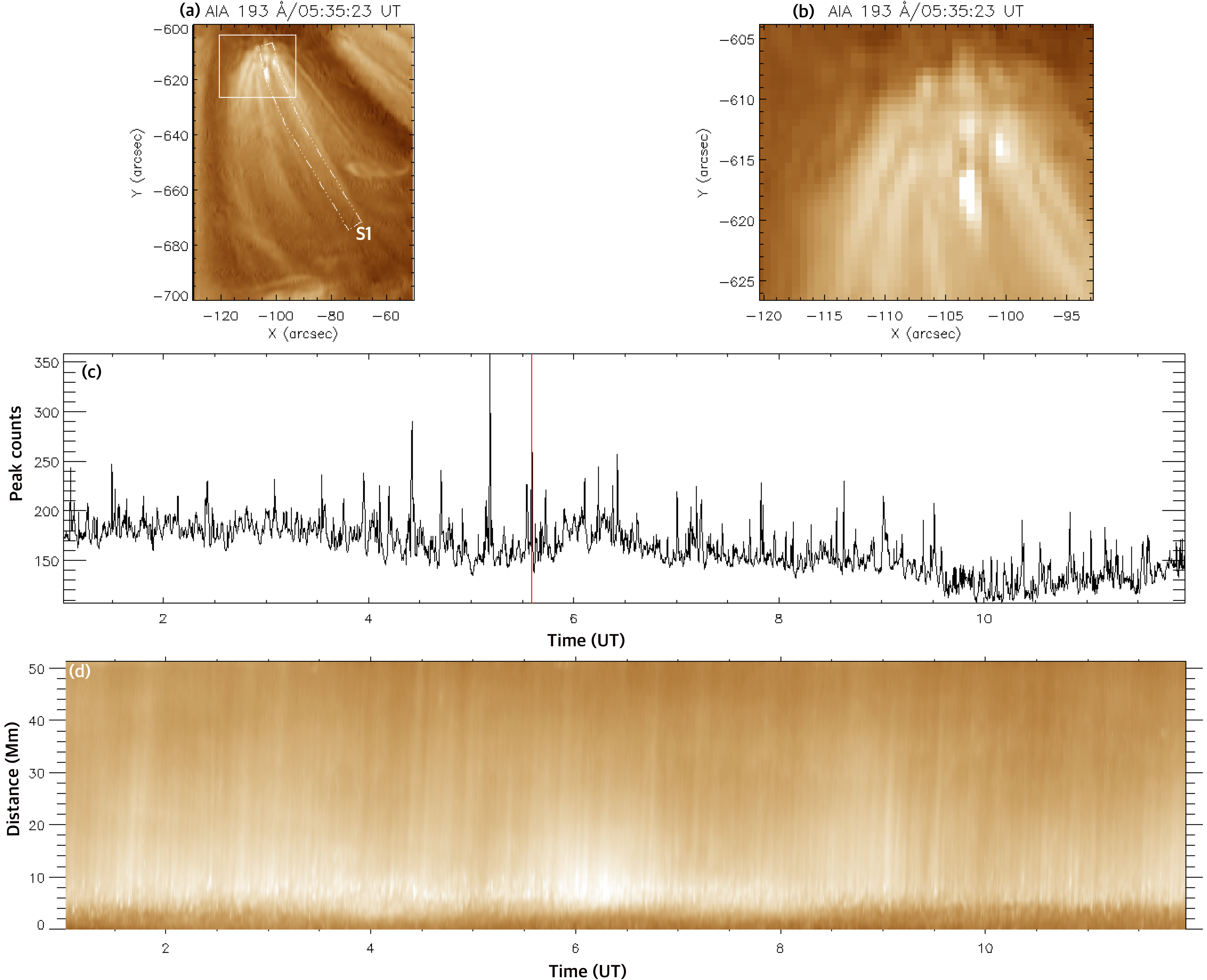}
}
\caption{(a,b) AIA 193 {\AA} images of the plume and zoomed view of the base (white box in (a)). (c) AIA 193 {\AA} peak counts (arbitrary units) extracted from the box region in (a).  (d) Time-distance intensity plot along slice S1 (shown in (a)) demonstrating jetlets and associated brightenings. An animation of panels (a-c) is available. The animation runs from 01:01:35 UT to 11:57:23 UT (November 20). } 
\label{fig6}
\end{figure*}
%%%%%%%%%%%%%%%%%%%%%%%%%%%%%%%%%%%%%%%%%%%%%%%%%%%%%%%%%%%%%%%

%%%%%%%%%%%%%%%%%%%%%%%%%%%%%%%%%%%%%%%%%%%%%%%%%%%%%%%%%%%%%%
\begin{figure*}
\centering{
\includegraphics[width=18cm]{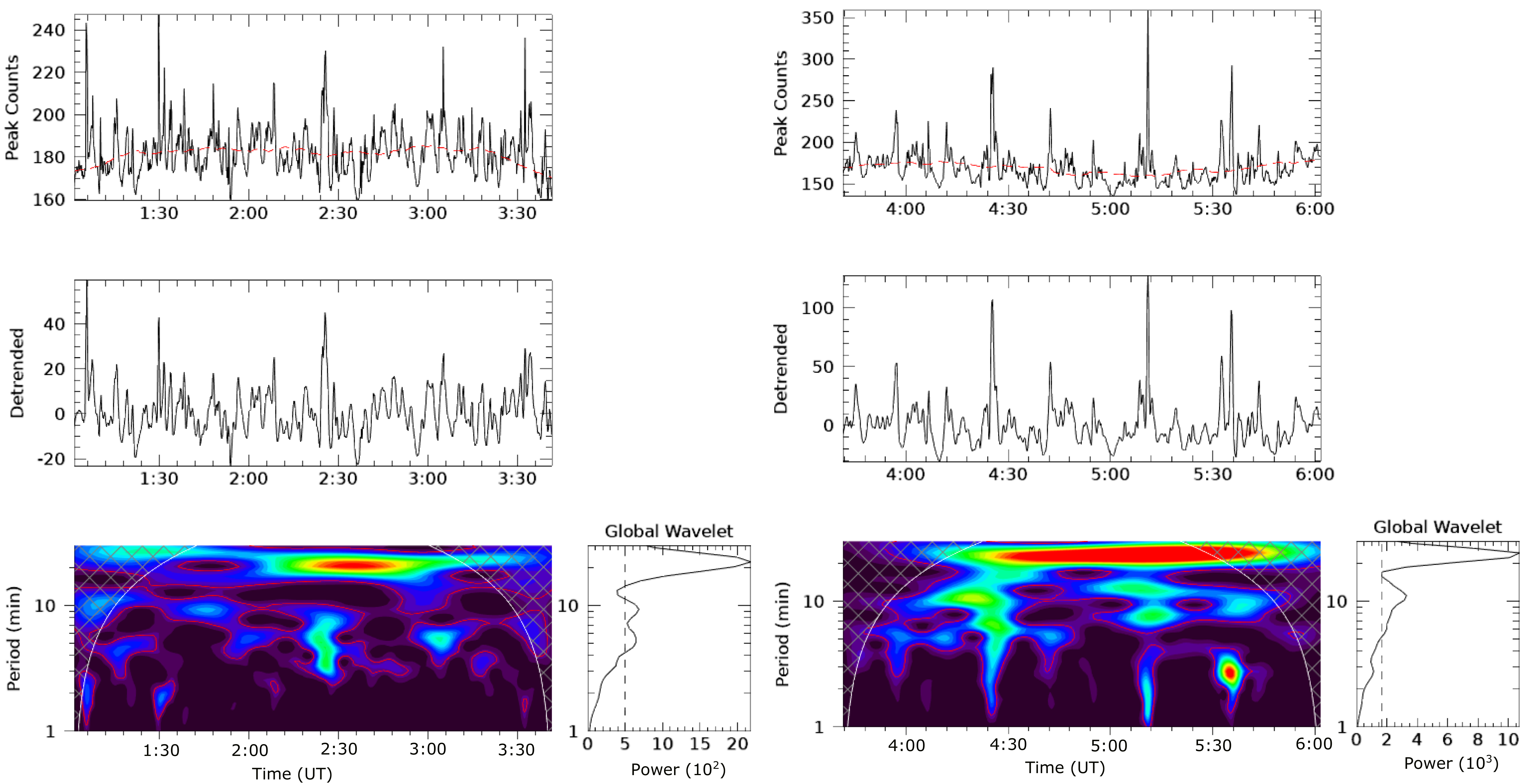}
\includegraphics[width=18cm]{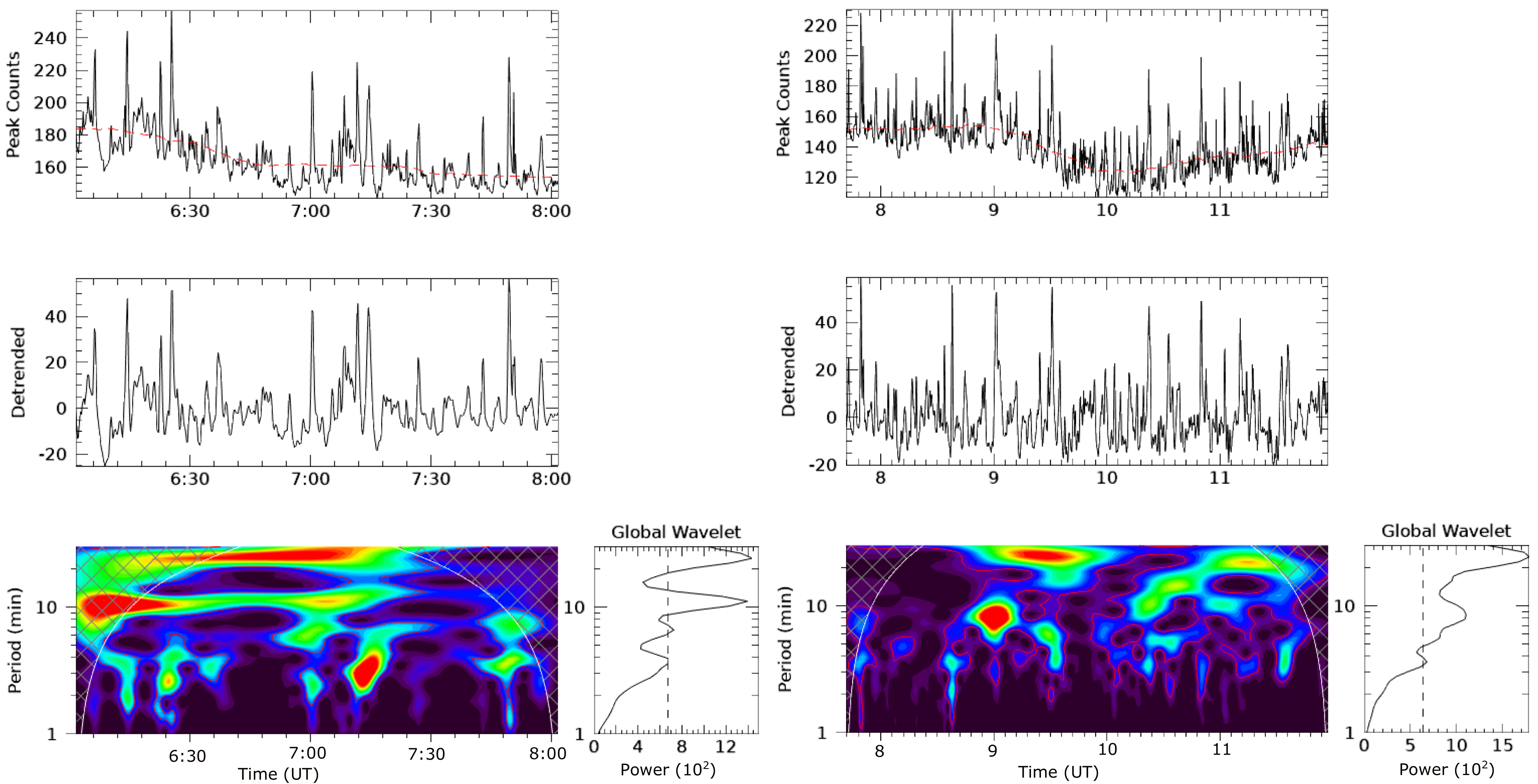}
}
\caption{AIA 193 {\AA} light curves and wavelet plots for the region shown in panel (b) for four intervals on November 20, using the same methods as in Figure \ref{fig3}).} 
\label{fig7}
\end{figure*}
%%%%%%%%%%%%%%%%%%%%%%%%%%%%%%%%%%%%%%%%%%%%%%%%%%%%%%%%%%%%%%%

\section{OBSERVATIONS AND RESULTS}\label{obs}
We analyzed SDO/AIA \citep{lemen2012} full-disk images of the Sun (field-of-view $\approx$ 1.3~$R_\sun$) with a spatial resolution of 1.5$\arcsec$ (0.6$\arcsec$~pixel$^{-1}$) and a cadence of 12~s, in the following channels: 304~\AA\ (\ion{He}{2}, temperature $T\approx 0.05$~MK), 171~\AA\ (\ion{Fe}{9}, $T\approx 0.7$~MK), and 193~\AA\ (\ion{Fe}{12}, \ion{Fe}{24}, $T\approx  1.2$~MK and $\approx 20$~MK) images. We used cotemporal
SDO/HMI \citep{schou2012} magnetograms to investigate any photospheric magnetic field changes during the jetlets. The 3D noise-gating technique \citep{deforest2017} was applied to clean the SDO/AIA and HMI images. 
To determine the underlying magnetic topology for jet source region, we utilized a potential-field extrapolation code \citep{nakagawa1972} available in the GX simulator package of SSWIDL \citep{nita2015}. 
We analysed PSP \citep{fox2016} solar wind and energetic particle observations from Electromagnetic Fields Investigation (FIELDS, \citealt{bale2016}), Solar Wind Electron Alpha Proton (SWEAP, \citealt{kasper2016}), and Integrated Science Investigation of the Sun (IS$\sun$IS, \citealt{mccomas2016}) instruments. 

During each approach of PSP to its orbital perihelion, there are two intervals of co-rotation (i.e., the spacecraft essentially dives toward a nearly fixed point on the Sun's surface) separated by an interval around perihelion in which it transits a significant distance across the solar surface. During Encounter 5 studied by \citet{fargette2021}, PSP was transiting, so its magnetic connection to the Sun mapped to a large range of longitudes. Those investigators converted the wide range of coordinates traversed by PSP during that time to a smaller range of coordinates at the Sun and analyzed the spatial distribution of the fluctuations observed. 
In contrast, on November 19-21 during Encounter 10, PSP was co-rotating while inbound \citep{bale2022,badman2023}, effectively hovering over a coronal hole while orbiting around the Sun. Therefore, the interval analyzed in our study was much longer than the duration of a single jetlet or jet (minutes), in fact, longer than several repeats of jetlet activity. We measured PSP signatures of quasiperiodic microstreams during the co-rotation phase and compared them to the periodicities of jetlets that occurred in the low solar atmosphere and were magnetically connected to PSP during this phase.

During encounter 10, PSP observed solar wind emanating from two equatorial CHs during 2021 November 19-21 \citep{bale2022}. The positions of multiple spacecraft including PSP on November 20 are displayed in Figure \ref{fig1}(a) using the Solar-MACH tool \citep{Gieseler2022}. The PSP position varies from 15-30 R$_\sun$(0.07-0.14 AU) during this encounter (Figure \ref{fig2}(d)). The Potential Field Source Surface (PFSS \citep{Altschuler1969,Schatten1969, Wang1992} extrapolation on November 20 reveals open fields (pink) emerging from CH1 (Figure \ref{fig1}(b)). SDO AIA 193 {\AA} and HMI magnetogram synoptic maps (Figure \ref{fig1}(c,d)) show the footpoint connectivity of PSP on 20 November \citep{Rouillard2020}. The PSP position, represented by a red X, was connected to CH1 during November 19-20 and CH2 on November 21.

Figure \ref{fig2}(a-c) displays the solar wind radial magnetic field (B$_R$), radial speed (V$_R$), and proton flux vs energy during 2021 November 19-21.  PSP detected fast-wind microstreams (V$_R$$>$500 \kms), associated switchbacks (B$_R$), and enhancements in the proton flux during this interval. More details about this encounter are described in \citet{bale2022}. IS$\sun$IS Energetic Particle Instrument-Low (EPI-Lo \citealt{hill2020}) observes solar energetic particles (SEPs) from 20 keV nucleon$^{-1}$ to 1.5 MeV $^{-1}$ over 2$\pi$ sr. The SEP ion composition is often utilized as a good indicator of the origin of SEPs \citep{reames2021}.
The SEP ion composition observations from EOI-Lo reveal multi-day enhancement in proton (H), helium ($^3$He, $^4$He), and heavy ions such as oxygen (O) and iron (Fe) during the microstreams/switchbacks (Figure \ref{fig2}(d-h)).
Note that there is a weak enhancement in $^3$He during the whole interval. The 
GOES soft X-ray flux profile in 1-8 {\AA} channel shows no flare activity during this microstream interval (Figure \ref{fig2}(i)). 

To investigate microstream periodicity, we used several subintervals from November 19 17:00 UT to November 21 14:00 UT. We performed a Morlet wavelet analysis \citep{torrence1998}  of the solar wind radial speed (V$_R$) in each sub-interval. To obtain a detrended light curve for each interval, the speed profile was first smoothed with a boxcar average of width 12 s, using the IDL SMOOTH routine, then a smoothed profile boxcar-averaged at 30 min was subtracted.
The wavelet analyses of these subintervals are shown in Figures \ref{fig3}, \ref{fig4}, and Appendix Figures \ref{figa1}, \ref{figa2}, and \ref{figa3}. These plots clearly demonstrate the presence of multiple periods close to 3, 5, 10, and 20 minutes in microstreams with above 99$\%$ significance level. 
The periods in all microstream subintervals are summarized in Table \ref{tab1}. Note that intervals with the same number on different days are not the same interval.

%%%TABLE%%%%%%%%%%%%%%%%%%%%%%%%%%%%%%%%%%%%%%%%%%%%%
{\small
\begin{longtable*}{c c c c c c c c c}
\caption{Periods (minutes) in V$_R$ at several intervals} \\
\hline \\
\label{tab1}
Date   &Interval \#1              &\#2     &\#3         &\#4        &\#5    &\#6       &\#7   &\#8  \\
(2021)  &              &     &         &        &     &        &    &   \\
           \hline
 Nov. 19  &2, 4, 8                 &3, 5, 7, 18     &2, 4, 6, 18         & -       &-     &-        &-    &-   \\          
 Nov. 20  &3, 5, 8, 15, 24     &2, 3, 8, 12     &3, 5, 8, 20    &2, 3.5, 7, 28   &2, 5, 8, 20     &3, 5, 8, 16         &-    &-   \\  
 Nov. 21  &2, 4, 20, 25          &3, 7, 18, 24  &2, 5, 8, 28     &3, 5, 9, 19      &4, 8, 15, 30   &2.5, 5, 10, 30   &5, 20    &5.5, 8.5, 13, 20, 28   \\            
                      
\hline
\end{longtable*}
\small
\noindent
}
%%%%%%%%%%%%%%%%%%%%%%%%%%%%%%%%%%%%%%%%
To determine the source region of these solar wind microstreams and low energy SEPs, we analysed PSP and SDO data on November 20, $\approx$1-12 UT, while PSP was connected to CH1 (Figures \ref{figa4}(a,b)). 
At this time CH1 contained several bright points and bright and faint plumes everywhere. PSP was connected to CH2 on November 21, which went behind the limb from the AIA point of view. Therefore, it was not possible to analyse the jet activities inside CH2.

The HMI magnetogram and accompanying animation show the presence of positive (minority) polarity everywhere in the negative (majority) polarity CH1 (Figure \ref{figa4}(d)). The photospheric magnetic field is saturated at $\pm$10 G above the noise level. Interestingly, the AIA 171 {\AA} animation shows quasiperiodic jets everywhere in the CH1 from November 19 20:12:45 UT to November 20 22:41:15 UT (Figure \ref{figa4}(c), marked by arrows). Microstreams and low-energy SEPs were detected by PSP during the same interval. One example of repetitive jetting from a CH1 bright point is shown in Figure \ref{figa4}(e-g)). The potential field extrapolation of the source region shows a classic fan-spine topology (Figure \ref{figa4}(h)). Quasiperiodic jets emanate from the cusp along the outer spine (period $\approx$10 and 20 minutes, Figure \ref{figa5}(d,e)), while the underlying minority-polarity patches interact with adjacent positive polarity and decay. The AIA 171 {\AA} time-distance intensity plot (Figure \ref{figa5}(c)) and accompanying animation along the rectangular slit S clearly reveal the quasiperiodic jets along the outer spine of this small fan-spine topology during 1-7 UT on November 20. The recurrent jets are associated with quasiperiodic brightenings (Figure \ref{figa5}(d)) in the fan.

We analyzed AIA 304, 171, and 193 {\AA} images of the jetlets at the base of the brightest plume in CH1 (marked by an arrow in Figure \ref{figa4}(a)) and larger jets scattered throughout the CH. The accompanying animation reveals quasiperiodic brightenings and associated jetlets (marked by arrows) randomly distributed at the base of the plume (Figure \ref{fig5}). The jetlets are associated with chromospheric/transition region ejections and brightenings in the 304 {\AA} channel that are $\approx$2-3$\arcsec$ wide. In addition, outflows (jetlets) propagated along the plumelets associated with these brightenings. 
 To estimate the periodicities of these energy-release episodes at the plume base, we extracted peak counts (arbitrary units) within a box (Figure \ref{fig6}(a)) at the footpoint of the plume during 1-12 UT on November 20 (Figure \ref{fig6}(a-c)). The quasiperiodic bursts in the light curve (see accompanying animation) are associated with jetlets and brightenings (Figure \ref{fig6}(d)), most likely generated by interchange reconnection near the base of the plume (e.g., \citealt{kumar2022}). The wavelet analysis reveals the presence of $\approx$3, 5, 10, and 20 minute periods (99$\%$ significance) in the emissions of the brightenings and jetlets (Figure \ref{fig7}). 

A low-energy SEP enhancement in proton flux and $^3$He, $^4$He, Fe, and O ion compositions was observed by PSP during the November interval discussed here. The field lines linked to PSP were well-connected to footpoints in CH1 and CH2 \citep{bale2022}, although the inherent uncertainties in tracing magnetic connectivity from the heliosphere to the photosphere prevent identifying these footpoints with specific features.  In addition, the GOES soft X-ray data reveal no flare activity during 19-21 November. Therefore we conclude that CH1 and CH2 are the sources of the SEP enhancement.

\section{DISCUSSION}\label{discussion}
We found periodicities in the microstreams observed by PSP during encounter 10 (15-30 R$_\odot$) on 2021 November 19-21, and compared them with AIA observations of impulsive events in the magnetically connected equatorial coronal holes. The estimated microstream periods are approximately 3, 5, 10, and 20 minutes. On November 19-20 PSP was connected to CH1, located on the disk, so we could identify and characterize signatures of energy release inside the CH using AIA observations. Several bright points and plumes were observed everywhere in the equatorial CH; HMI magnetograms show that these features were associated with positive-polarity sites scattered throughout the negative polarity CH. PSP observations also revealed low-energy multiday SEP enhancements while magnetically connected to CH1.

Quasiperiodic jets on a range of spatial scales emanated from the plume bases and from those bright points. Note that jetlets manifesting similar periodicities have been observed at the base of other plumes in four different equatorial CHs \citep{kumar2022}. We found that the jetlets are heated to 1-2 MK and attain speeds of order 100 \kms; each impulsive event was estimated to release approximately 10$^{24}$ erg. Recurrent jets with periods of $\approx$6-10 minutes also have been observed in small bright points in other coronal holes \citep[e.g., ][see animation \href{https://zenodo.org/record/2555389/files/jet26A.mp4?download=1}{1}]{kumar2019a,kumar2022}. Periods of 5-30 minutes have also been observed in outflows from polar plumes \citep{mcIntosh2010,tian2011}. The present analysis shows that the microstream periods agree well with the periods of the repetitive EUV brightenings and jetlets at plume bases. 

In addition, PSP recorded an extended interval of enhanced $^3$He and heavier ions that coincided with the interval containing the multiperiodic microstreams and jetlets. Ulysses observations showed that microstreams are associated with variations in the helium content (i.e., ratio of alpha particle flux to the proton flux) of the solar wind, indicating their solar origin \citep{neugebauer1997}. For PSP encounter 10 specifically, \citet{bale2022} reported fluctuations in the helium content similar to the microstream variations, and suggested the chromospheric origin of the microstreams. Multiday enhancements in $^3$He, $^4$He, Fe, and O were detected by ACE at 1 AU on October 12–16, 2005, attributed to AR 10814 \citep{wang2006}, but it was unclear why the $^3$He enhancement continued for several days. \citet{mason2007} suggested that continuous magnetic reconnection would release $^3$He into the interplanetary medium, but the exact source region was not identified due to insufficient resolution of the available EUV observations. Note that big jets are often associated with impulsive $^3$He-rich events at 1 AU \citep{wang2006,pick2006,mason2016} along with enhancements in heavy ions (Fe, O etc.). Because the same physical mechanism is responsible for jets on all scales, we conclude that the full panoply of jets, including jetlets, is likely to produce similar ion enhancements. The coexistence of quasiperiodic outflows from numerous positions within CH1, microstreams with the same periods detected at PSP, and enhanced ionic composition at PSP consistent with coronal jets strongly suggests a direct connection between jetlets at the Sun and microstreams in the heliosphere. 

Our prior numerical and observational studies of coronal jets provide important clues about the underlying physics linking jetlets and microstreams. Both jets and jetlets emanate from embedded bipoles, which have classic fan-spine topologies \citep{antiochos1998}. When stressed by emergence and/or footpoint motions, the null atop the fan is distorted into a current sheet, enabling reconnection between the internal (closed) magnetic flux inside the separatrix and the external (open, in CHs) flux. This interchange reconnection alone tends to produce weak jets \citep{pariat2009,pariat2015,kumar2021}, whereas the presence of a filament channel inside the fan can yield more energetic and explosive outflows according to the breakout jet model \citep{wyper2017,wyper2018a}. Our earlier jetlet study found that the fan surface is $\approx$2-3 arcsec wide and the height of the null point lies in the chromosphere/transition-region \citep{kumar2022}. Although most jetlet sources are too small to allow detection of mini-filaments, nearly all larger coronal jets exhibit this and/or other key signatures of filament channels \citep{sterling2015,kumar2019b}. Therefore, we propose that repetitive reconnection in multiple locations beneath plumes and elsewhere in CHs is responsible for initiating and driving outflows that evolve into microstreams. 

Our previous and present plume observations show that the jetlet widths lie between 1-2 arcsecs (750-1500 km), while plumes are 30-50 arcsecs (22500-37500 km) wide. Therefore, the scale sizes of individual microstreams and patches projected back to the Sun by previous studies \citep{bale2021,fargette2021,bale2022} are roughly consistent with the sizes of jetlets (granules) and plumes (supergranules). 
We have also demonstrated that mixed polarities and ubiquitous jetlets/jets from plumes and bright points exist throughout CHs. When PSP co-rotates over a CH, it is magnetically connected to these solar source regions for intervals substantially longer than the duration of individual intensity variations. The recurrent patterns found in jetlets/jets are mirrored in the microstreams, implying that they have a common origin (i.e., interchange reconnection). The discovery of jetlet periods commensurate with p-modes raised fundamental questions about this relationship \citep{uritsky2021,kumar2022}. Two possibilities exist: reconnection could be induced by p-mode waves impinging on the null \citep{chen2006,heggland2009}, or the observed flows could be modulated by p-modes. Future numerical simulations will aid in determining which explanation is more likely.

\section{CONCLUSIONS}\label{conclusion} 

We have demonstrated, for the first time, that periodicities in the radial velocities of microstreams are consistent with the periodicities observed in the EUV emissions of recurrent jetlets at the base of plumes and throughout coronal holes. This evidence, coupled with the coincident enhancement of $^3$He and heavier ions, suggests a direct link between jetlets and microstreams, including the associated magnetic switchbacks. A significant gap remains, however: we have traced jetlets outward into plumelets, but individual parcels of plasma have not been followed from plumes to the inner heliosphere. Whether these narrow, high-speed injections can transit this crucial zone and produce the characteristic magnetic-field reversals of switchbacks remains to be demonstrated. The Alfv\'enic fronts and wakes propagating well ahead of the dense jetlet material may play a role in forming the microstream/switchback properties. One approach to solving this problem is to simulate self-consistently the generation and transport of reconnection-driven jets into a more realistic solar wind than has been modeled thus far. 

Quasiperiodic jetlets (energy equivalent to nanoflares) can contribute to coronal heating and significant mass flux to the solar wind \citep{kumar2022}. 
The omnipresence of jetlets also may explain the origin of the fast solar wind in CHs \citep{raouafi2023}. We anticipate more detailed observations and analyses of similar events from the upcoming PSP and SolO \citep{muller2013} perihelions in the near future.  

%%%%%%%%%%%%%%%%%%%%%%%%%%%%%%%%%%%%%%%%%%%%%%%%%%%%%%%%%%%%%%%%%%%%
%\begin{acknowledgments} 
\acknowledgments
We thank the referee for insightful comments that have improved this paper. SDO is a mission for NASA's Living With a Star (LWS) program. PK thanks Tibebu Getachew Ayalew and Christina Cohen for their help with the ISI$\sun$S GUI package. PK thanks Michael Stevens and Justin Kasper for their help in accessing the PSP data. 
Parker Solar Probe was designed, built, and is now operated by the Johns Hopkins Applied Physics Laboratory as part of NASA’s Living with a Star (LWS) program (contract NNN06AA01C). Support from the LWS management and technical team has played a critical role in the success of the Parker Solar Probe mission. Thanks to the FIELDS team  (PI: Stuart D. Bale, UC Berkeley), the Integrated Science Investigation of the Sun (IS$_\sun$IS) Science Team (PI: David McComas, Princeton University), and the Solar Wind Electrons, Alphas, and Protons (SWEAP) team for providing data (PI: Justin Kasper, BWX Technologies). The Solar-MACH tool was originally developed at Kiel University, Germany and further discussed within the ESA Heliophysics Archives USer (HAUS) group.
Magnetic-field extrapolation was visualized with VAPOR (\href{(www.vapor.ucar.edu}{www.vapor.ucar.edu}), a product of the Computational Information Systems Laboratory at the National Center for Atmospheric Research.
 This research was supported by NASA's Heliophysics Guest Investigator (\#80NSSC20K0265) and GSFC Internal Scientist Funding Model (H-ISFM) programs, and the NSF SHINE program (Award Number \#2229336). Wavelet software was provided by
C. Torrence and G. Compo, and is available at \href{http://paos.colorado.edu/research/wavelets/}{http://paos.colorado.edu/research/wavelets/}.
%\end{acknowledgments}
%%%%%%%%%%%%%%%%%%%%%%%%%%%%%%%%%%%%%%%%%%%%%%%%%%%%%%%%%%%%%%%%%%%%
%%%%%%%%%%%%%%%%%%%%%%%%%%%%%%%%%%%%%%%%%%%%%%%%%%%%%%%%%%%%%%%%%%%%

\bibliographystyle{aasjournal}
\bibliography{reference.bib}

%%%%%%%%%%%%%%%%%%%%%%%%%%%%%%%%%%%%%%%%%%%%%%%%%%%%
\appendix
\counterwithin{figure}{section}
\section{Microstream periodicity analysis and CH jets}
This section contains supplementary material (additional figures) to support the results described above. The high-quality animations for Figures 5, 6, A4, and A5 are available in the Zenodo repository at doi: \href{https://zenodo.org/record/7924233\#.ZFy8fezMJ58}{10.5281/zenodo.7924233}.
%%%%%%%%%%%%%%%%%%%%%%%%%%%%%%%%%%%%%%%%%%%%%%%%%%%%%%%%%%%%%%
%\setcounter{figure}{0}
\begin{figure*}[htp]
\centering{
\includegraphics[width=18cm]{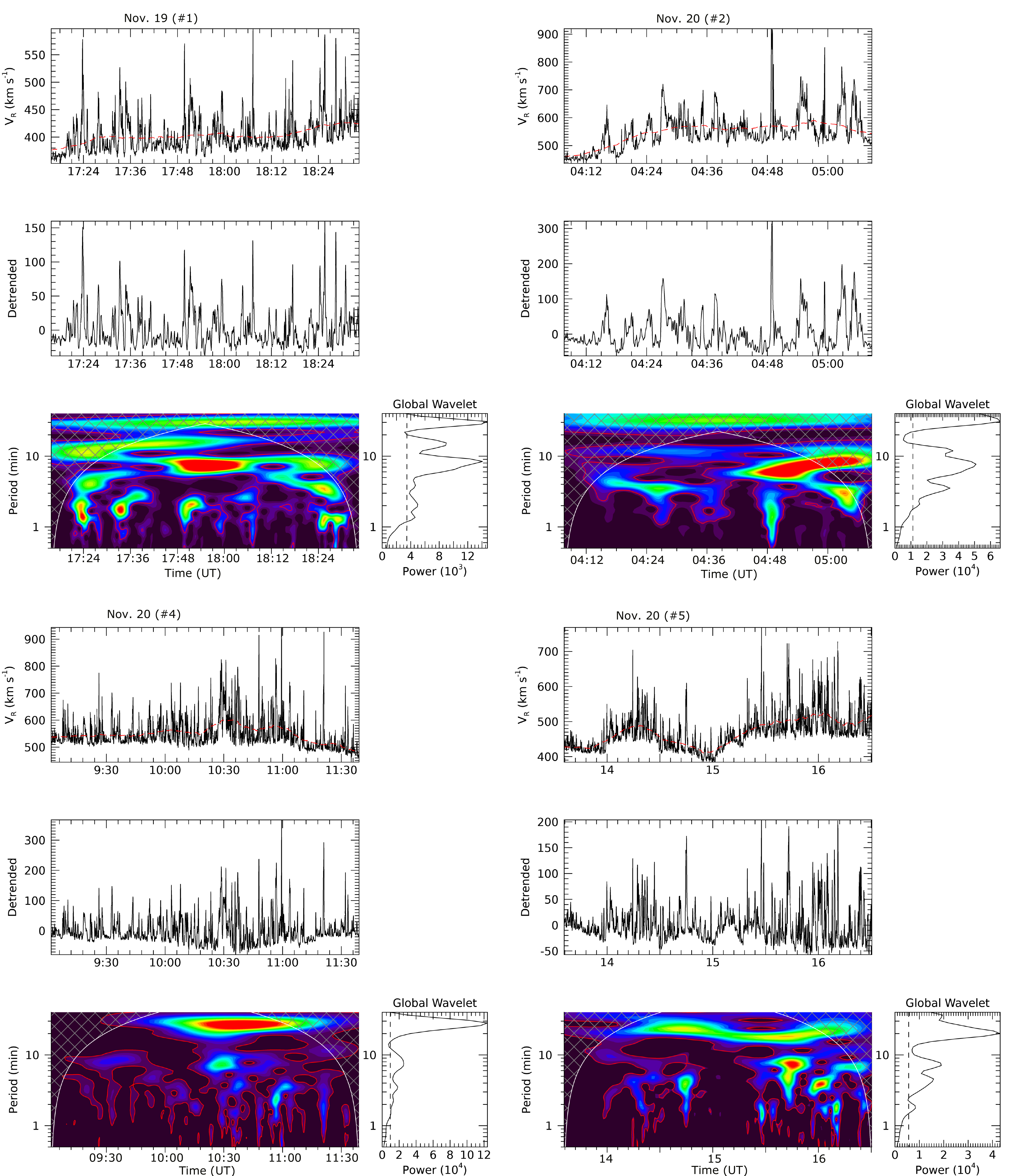}
}
\caption{Microstream periodicities in Interval \#1 on 2021 November 19 and Intervals \#2, 4, and 5 on 2021 November 20, using the same methods as in Figure \ref{fig3}.}
\label{figa1}
\end{figure*}
%%%%%%%%%%%%%%%%%%%%%%%%%%%%%%%%%%%%%%%%%%%%%%%%%%%%%%%%%%%%%%%
%%%%%%%%%%%%%%%%%%%%%%%%%%%%%%%%%%%%%%%%%%%%%%%%%%%%%%%%%%%%%%
\begin{figure*}[htp]
\centering{
\includegraphics[width=18cm]{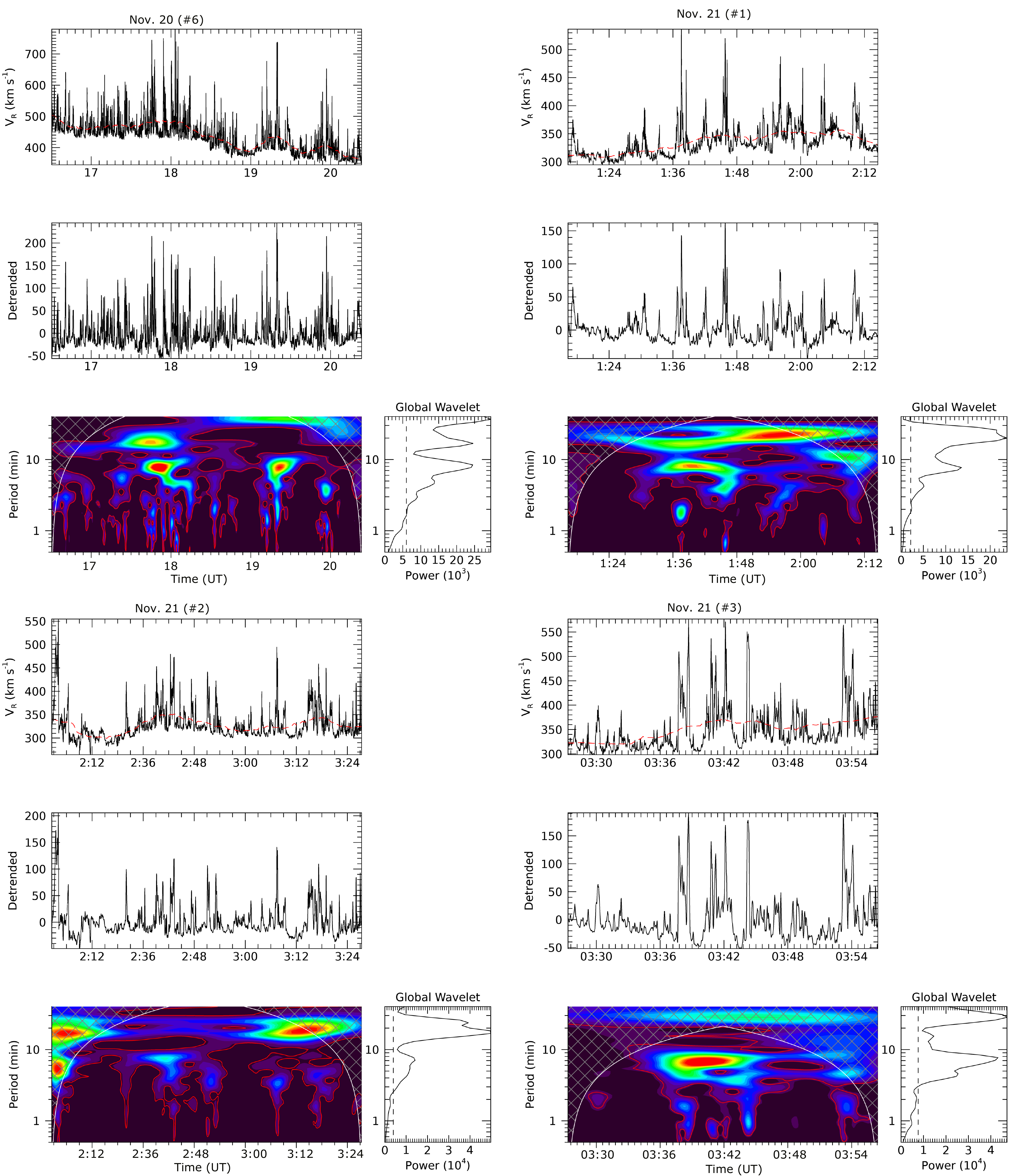}
}
\caption{Microstream periodicities in Interval \#6 on 2021 November 20 and Intervals \#1-\#3 on November 21, using the same methods as in Figure \ref{fig3}.}
\label{figa2}
\end{figure*}
%%%%%%%%%%%%%%%%%%%%%%%%%%%%%%%%%%%%%%%%%%%%%%%%%%%%%%%%%%%%%%%
%%%%%%%%%%%%%%%%%%%%%%%%%%%%%%%%%%%%%%%%%%%%%%%%%%%%%%%%%%%%%%
\begin{figure*}[htp]
\centering{
\includegraphics[width=18cm]{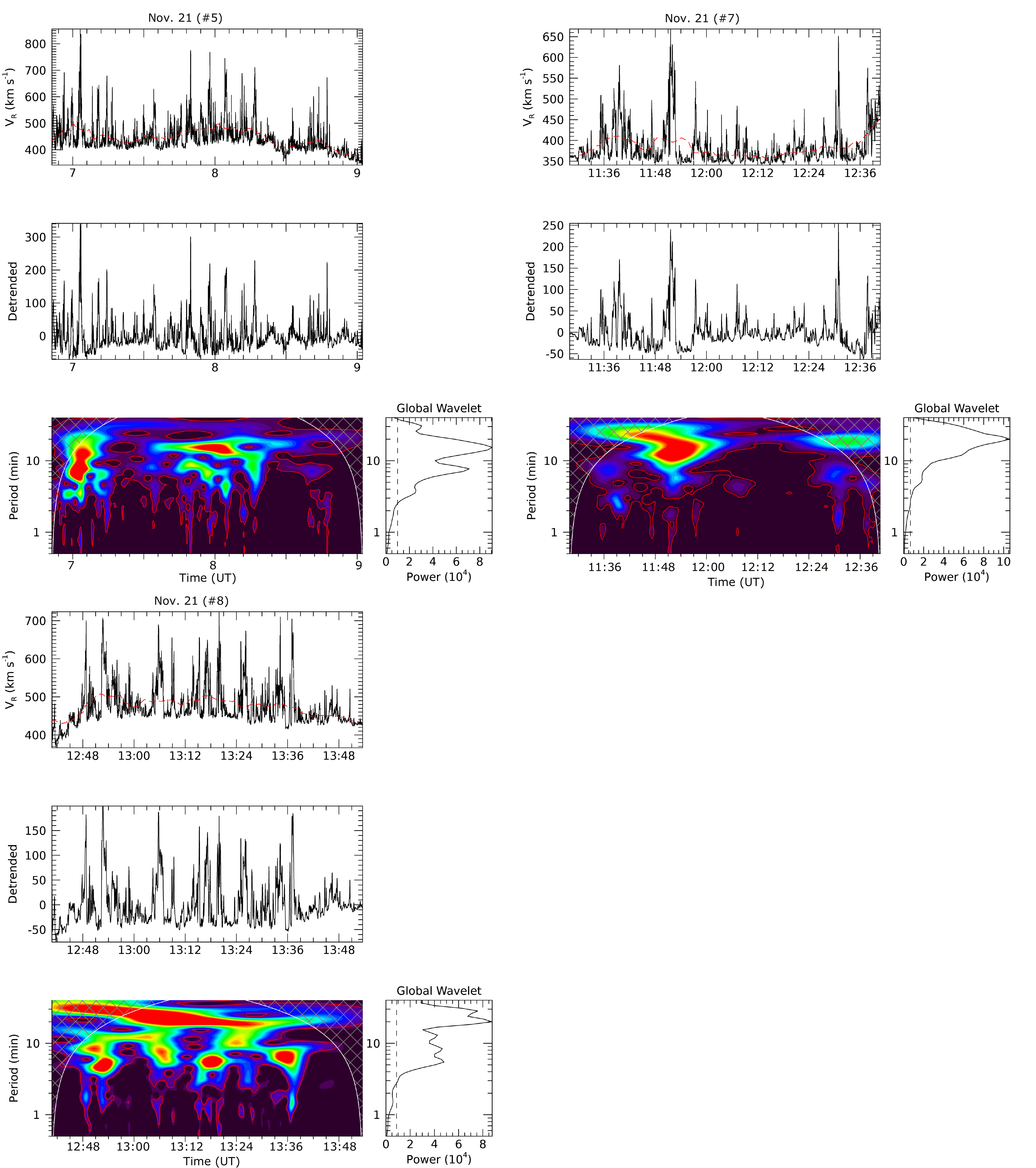}
}
\caption{Microstream periodicities in Intervals \#5, 7, and 8 on 2021 November 21, using the same methods as in Figure \ref{fig3}.} 
\label{figa3}
\end{figure*}
%%%%%%%%%%%%%%%%%%%%%%%%%%%%%%%%%%%%%%%%%%%%%%%%%%%%%%%%%%%%%%%
%%%%%%%%%%%%%%%%%%%%%%%%%%%%%%%%%%%%%%%%%%%%%%%%%%%%%%%%%%%%%%
\begin{figure*}[htp]
\centering{
\includegraphics[width=0.95\textwidth]{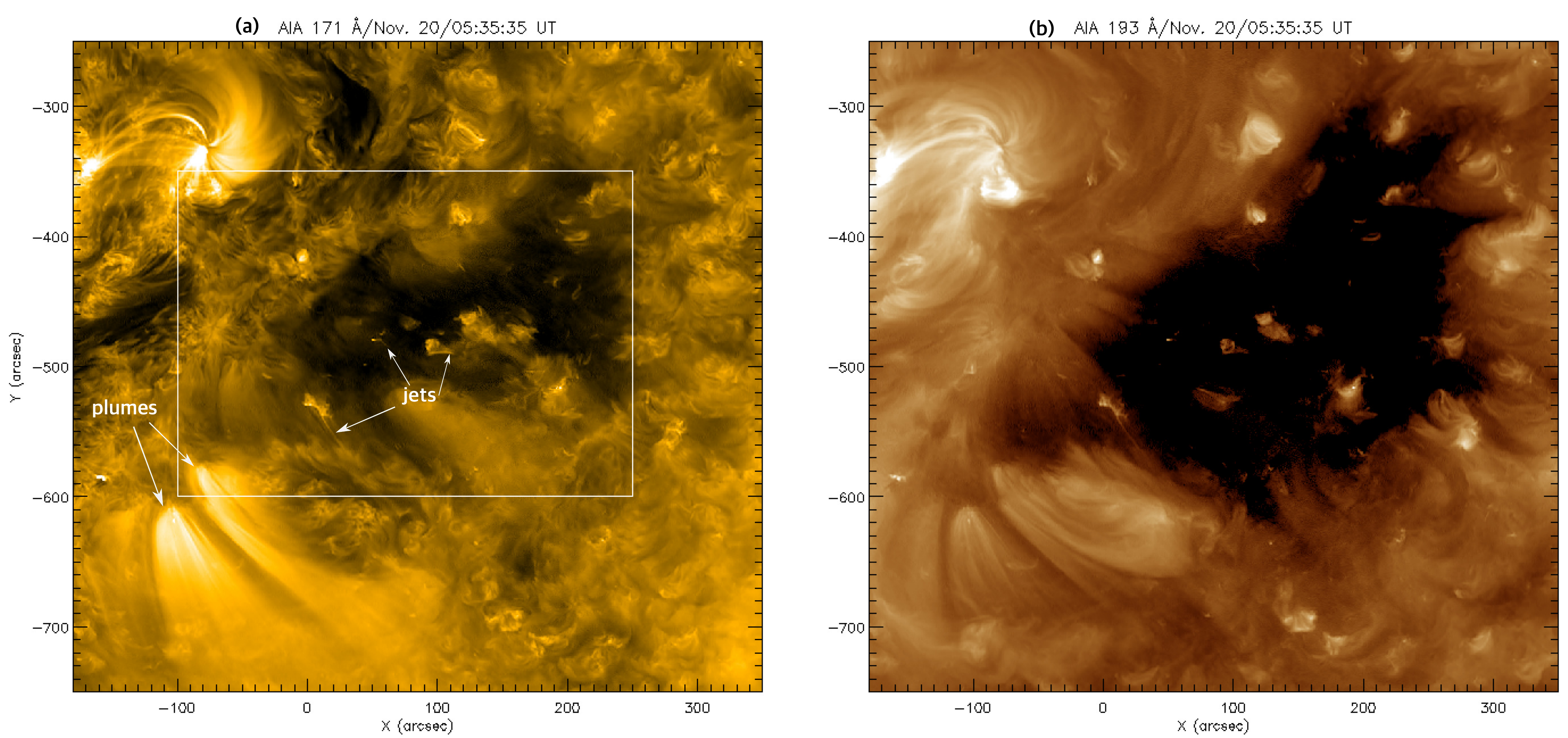}
\includegraphics[width=0.95\textwidth]{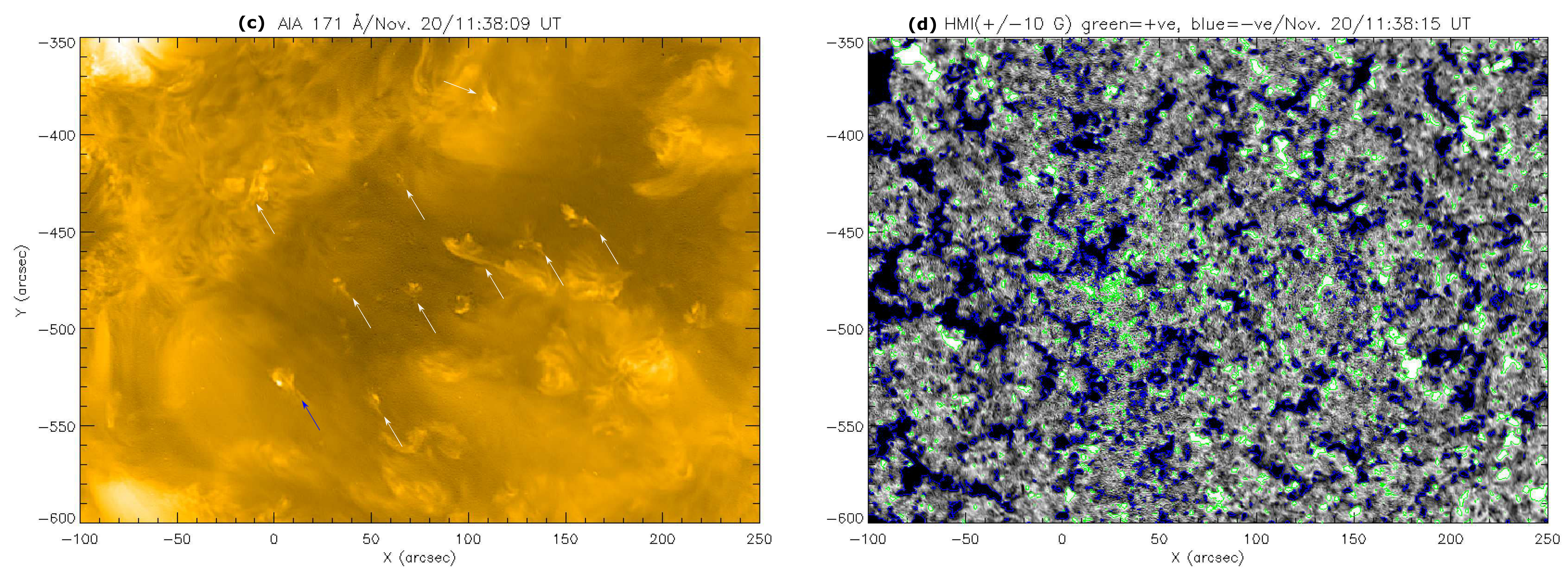}
\includegraphics[width=0.95\textwidth]{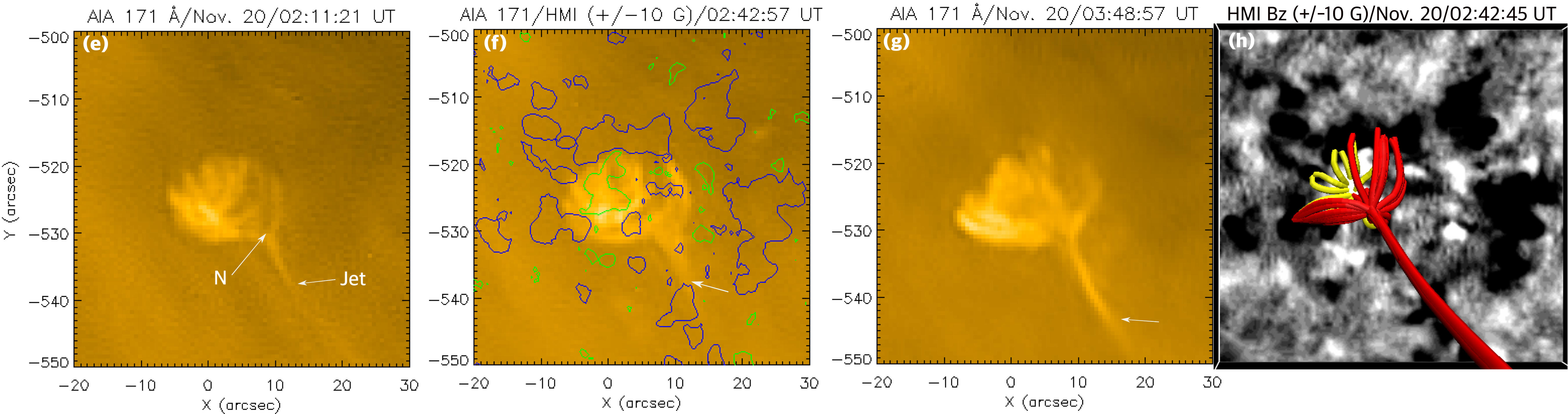}
}
\caption{(a,b) AIA 171 and 193 {\AA} images of CH1 on 2021 November 20. (c) Zoomed view of the white box in (a). Jets are marked by arrows. (d) HMI magnetogram (B=$\pm$10 G) of the white box in (a). Green (blue) is positive (negative) polarity. (e,g) AIA 171 {\AA} image of recurrent jets (marked by arrow) from the bright point in CH1 marked by a blue arrow in (c). N=null-point. (f) Same image as in (e) overlaid by HMI LOS field-strength contours (B=$\pm$10 G). Jet is marked by an arrow.  (h) Potential field extrapolation of the jet source showing the fan-spine magnetic topology. Yellow (red) field lines are closed (open). 
An animation of this Figure is available. The first part of the animation shows AIA 171 {\AA} and HMI magnetogram images with a larger field of view (FOV). The second part shows AIA 171 {\AA} and magnetogram images of a smaller FOV. The third part shows HMI magnetogram contours ($\pm$10 G) overlaid on 171 {\AA} images. The fourth part shows AIA 171 {\AA} images of recurrent jets from the bright point in CH1 marked by a blue arrow, overlaid by HMI magnetogram contours ($\pm$10 G). All animations (\#1-3) run from November 19 20:12:45 UT to November 20 22:41:15 UT. Animation \#4 runs from 01:00:45 UT to 12:05:09 UT on November 20. }
%The high-quality animations of this Figure are available in the Zenodo repository at doi: \href{https://zenodo.org/record/7924233\#.ZFy8fezMJ58}{10.5281/zenodo.7924233}.}
%The animations (\#1-3, size$>$100 MB) are available here: \href{https://drive.google.com/file/d/1jsEIQhS1in-O4TOiRDbndht3HbpotUG9/view?usp=share_link}{https://drive.google.com}.}
\label{figa4}
\end{figure*}
%%%%%%%%%%%%%%%%%%%%%%%%%%%%%%%%%%%%%%%%%%%%%%%%%%%%%%%%%%%%%%%

%%%%%%%%%%%%%%%%%%%%%%%%%%%%%%%%%%%%%%%%%%%%%%%%%%%%%%%%%%%%%%
\begin{figure*}[htp]
\centering{
\includegraphics[width=13cm]{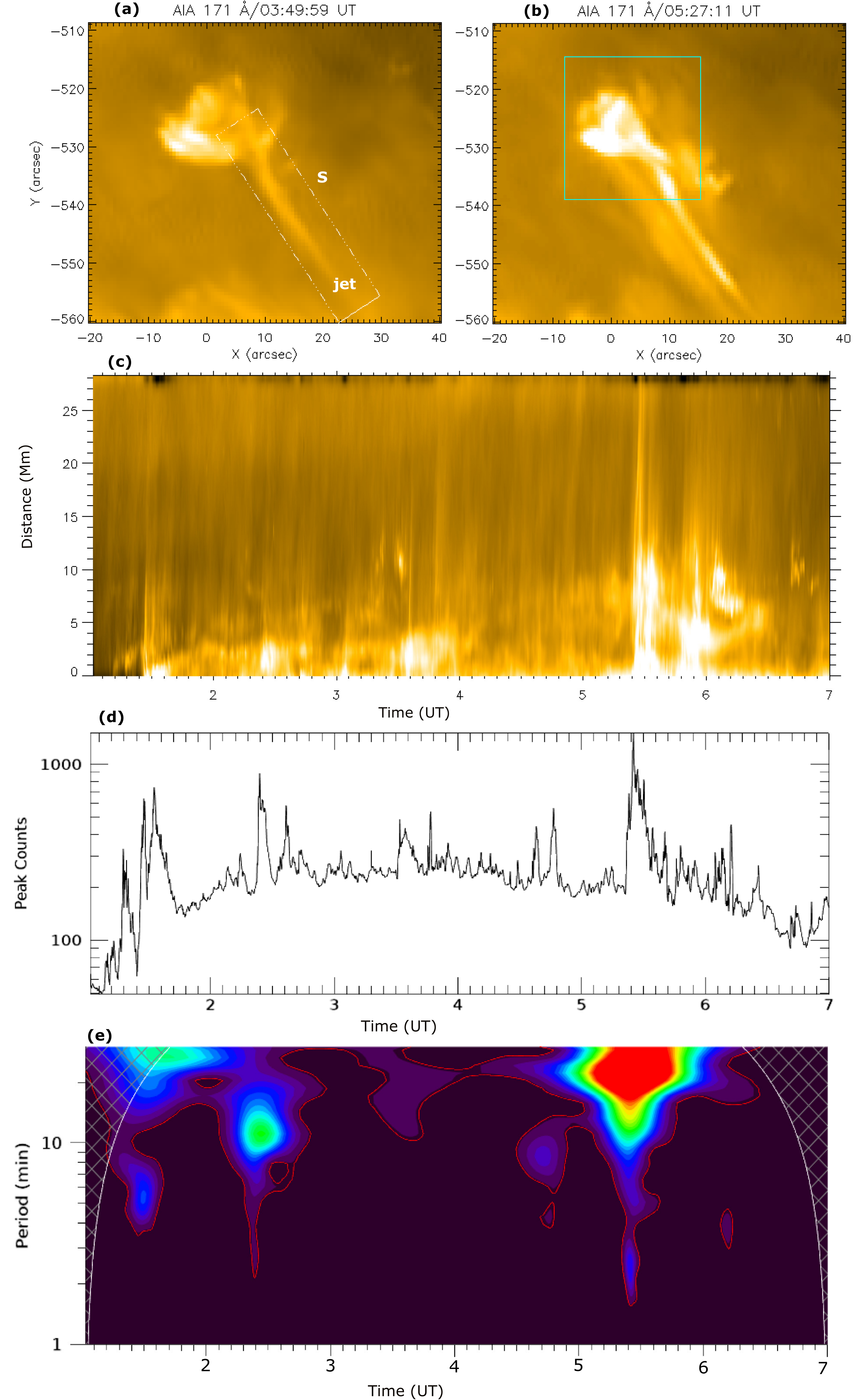}
}
\caption {AIA 171 {\AA} time-distance intensity plot (c) along the rectangular slice S (outer spine) in panel (a) showing quasiperiodic jets from a bright point. Peak counts (in the arbitrary unit, d) from the bright point using a box (cyan) marked in panel (b). (e) Wavelet power spectrum of the detrended signal. Red contours outline the 99\% significance level. The quasiperiodic intensity fluctuations (period$\approx$10, 20 min) are associated with jets from the bright point (tiny null-point topology). An animation of this Figure (a-c) is available. The animation runs from 01:00 UT to 07:00 UT on November 20.}
\label{figa5}
\end{figure*}
%%%%%%%%%%%%%%%%%%%%%%%%%%%%%%%%%%%%%%%%%%%%%%%%%%%%%%%%%%%%%%%

\end{document}